\documentclass[12pt,times]{article}
\usepackage{amsmath,amssymb,amsfonts}
\usepackage{algorithmic} 
\usepackage{graphicx}
\usepackage{textcomp}
\usepackage{xcolor}
\usepackage[square, comma, sort, numbers, compress]{natbib}
\usepackage{hyperref}
\hypersetup{
    colorlinks=true,
    linkcolor=black,
    filecolor=black,      
    urlcolor=brown,
    citecolor=black,
    }
\def\BibTeX{{\rm B\kern-.05em{\sc i\kern-.025em b}\kern-.08em
    T\kern-.1667em\lower.7ex\hbox{E}\kern-.125emX}}
\begin{document}

\title{Biomimetic Frontend for Differentiable Audio Processing\thanks{Supported by ONR Award N00014-23-1-2086 and a Dolby gift to UMD.}}
\author{Ruolan Leslie Famularo$^{1,2,3}$,  Dmitry N. Zotkin$^1$, \\
Shihab A. Shamma$^{4,3}$,  Ramani Duraiswami$^{1,3}$\\$^1$Perceptual Interfaces and Reality Lab., Computer Science \& UMIACS;\\  
$^2$Department of Linguistics;\\ 
$^3$Program in Neuroscience \& Cognitive Science; \\ 
$^4$Institute of Systems Research \& Electrical and Computer Engineering \\    
All at University of Maryland, College Park, USA. \\ 
\{rlli, wdz, sas, ramanid\}@umd.edu}
\date{September 12, 2024}
\maketitle
\begin{abstract} 
While models in audio and speech processing are becoming deeper and more end-to-end, they as a consequence need expensive training on large data, and are often brittle. We build on a classical model of human hearing and make it differentiable, so that we can combine traditional explainable biomimetic signal processing approaches with deep-learning frameworks. This allows us to arrive at an expressive and explainable model that is easily trained on modest amounts of data. We apply this model to audio processing tasks, including classification and enhancement. Results show that our differentiable model surpasses black-box approaches in terms of computational efficiency and robustness, even with little training data. We also discuss other potential applications.
\\
{\bf Keywords:} \ neuromorphic computing, bio-inspired AI, differentiable programming, auditory neuroscience, differentiable signal processing
\end{abstract}
\maketitle
\section{Introduction}
Audio and speech processing have recently moved towards deep and end-to-end architectures that learn from data, and achieved super-human performance on certain tasks. However, training such models is expensive in terms of both computation and data. For example, pretraining a speech model for downstream tasks requires hundreds of hours of data and access to expensive GPU clusters. 
These deep models lack robustness and are vulnerable to adversarial attacks, with drastic loss of performance from imperceptible modifications to inputs \cite{wu2022characterizing}. Also, these models are ``black-box'' and difficult to explain. 

In comparison, the human auditory system can effortlessly perform diverse tasks, including speech/speaker recognition and separation/enhancement. Classical audio and speech processing used features for machine learning that mimicked human perceptual processing, For example, the mel-frequency cepstral coefficient (MFCC) is conceptually similar to the cochlear analysis of sound (e.g., \cite{meyer2011robustness}). Although biomimetic methods are falling out of fashion due to the performance of pure input-output deep learning, we believe that integrating them into deep models may achieve the best of both worlds, allowing improved data/computational efficiency, robustness, and explainability. 

\textbf{Using Forward Models in Learning:}\ Historically, domain scientists and mathematicians have developed highly successful {\em forward mathematical models} connecting inputs to outputs. Such models are theory-driven, fast, and give actionable explanations, but usually perform poorly in the inverse settings where data has to be related to inputs. Deep learning models, however, allow learnable relationships between data and inputs via universal function approximations through optimization on appropriate cost functions.  Automatic differentiation (AD) is an under-appreciated but foundational element of this process. Derivatives collected from output to input allow ``backpropagation” by AD, a key ingredient of stochastic gradient descent used to train deep networks. Differentiable programming frameworks such as PyTorch and JAX make this almost invisible to the programmer when using standard network models. Differential Physics \cite{Thuerey2021-gf}, extends the concept of training based on a differentiable forward map to classical mathematical physics, allowing gradient-based optimization, and training of networks that respect physics and learn the parameters. We adapt this approach to auditory neuroscience forward models and show possible benefits: the known science is in the forward map; the approach works in sparse data and generative regimes; model fitting uses tremendous advances in hardware, ``tooling,'' and algorithms from deep learning. Related differentiable approaches have been applied to other fields, including Operator Learning (e.g., \cite{Li2020-tr}) and Neural Radiance fields (NeRFs) \cite{Xie2022-wo}.

In classical, non-differentiable settings, auditory models of the ear and brain \cite{mesgarani2006discrimination, elhilali2008cocktail, meyer2011robustness} have been applied to audio processing. These models outperformed traditional features such as MFCCs, especially in the presence of noise. Recently, several audio front-ends \cite{ravanelli2018speaker, zeghidour2021leaf} have been proposed to improve audio tasks in performance and/or robustness. Features based on cortical processing have also been combined with differentiable approaches to improve performance/robustness as a frontend \cite{vuong2020learnable} and as a loss function \cite{vuong2021modulation}. We seek to extend upon prior work and create a multi-stage differentiable model that adheres to biological auditory processing. 

\begin{figure*}[bht!]
\centering
\includegraphics[width=\textwidth]{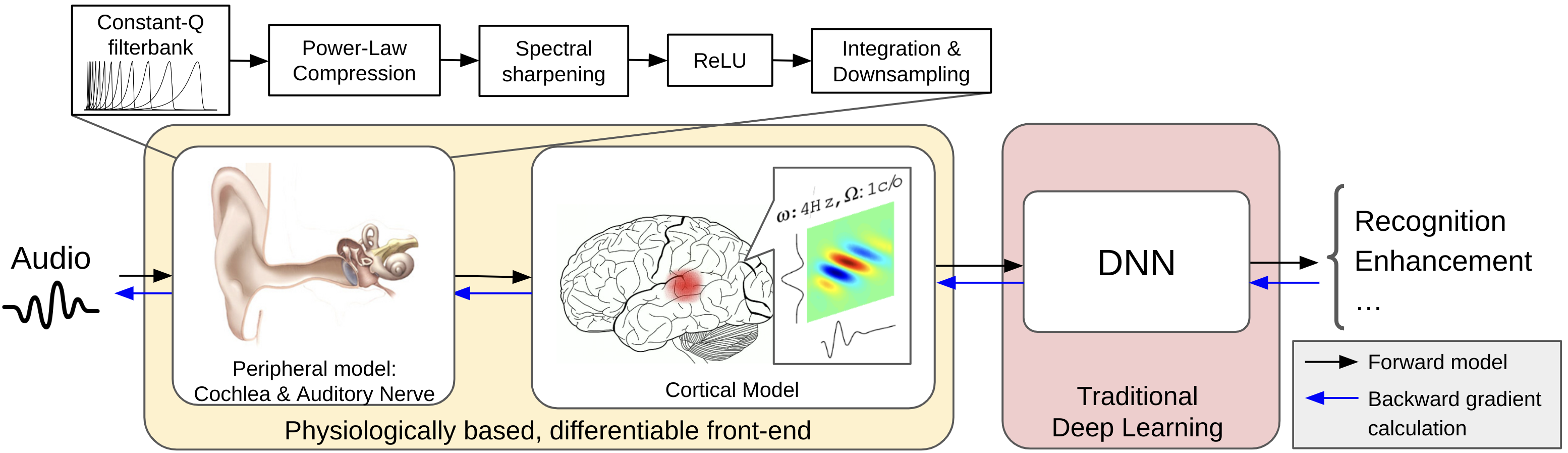}
\caption{Auditory processing from cochlea to cortical representations is shown from left to right. Black arrows indicate the forward model. Blue arrows indicate the direction of gradient calculation using the chain rule.}
\label{fig:model} 
\end{figure*}

We implement a differentiable and lightweight audio front-end informed by auditory neuroscience from the ear to the brain. It allows joint learning of cochlear and cortical parameters, as well as parameters of a backend neural network. We introduce the model and showcase example applications, highlighting the models' robustness and interpretability. To foster replicability and new applications, we made the code and parameters available at \href{https://github.com/pirl-lab/diffAudNeuro}{diffAudNeuro}.
\section{The auditory processing model}
We base our work on the forward model in \cite{Shamma1986-be, chi2005multiresolution, elhilali2004neural}, which is one of a host of models for peripheral auditory processing. 
These models vary considerably in the details of the various stages of filtering and transduction 
, with emphasis on cochlear mechanics \cite{Skrodzka2005-lu, de-Boer2000-wl}, hearing-aid design \cite{Moore1996-lq, Zhang2011-nv}, or simplifications towards efficient engineering applications \cite{ravanelli2018speaker, zeghidour2021leaf}.
The version we use provides a mathematically tractable model of early auditory processing that supplies all the essential details but remains mathematically tractable. 
\subsection{Step 1. A Model of the Cochlear and Peripheral Hearing}
The first stage performs computations of the auditory periphery, and converts input audio into a time-frequency representation, in four steps: a filterbank that performs frequency decomposition, nonlinear compression, lateral inhibition network (filter along frequency dimension), and lowpass pooling. 
The filterbank step consists of a bank of constant-$Q$ filters, corresponding to frequency decomposition in the cochlea. The filters were shaped to have a frequency domain response that approximates a rounded exponential (roex) function \cite{chi2005multiresolution, lyon2017human}: 
\begin{equation}
\text{if $x \in [0, x_h]$},\, |H(x)| = (x_h - x)^\alpha e^{-\beta (x_h - x)} \,
\text{ else, }   H(x)=0 
\end{equation}
where $x_h$ represents the highest frequency in the passband for one filter 
and $\alpha=0.3$ and $\beta=8$ are constants chosen to match the lower and upper skirts from neurobiological measurements \cite{allen1985cochlear}, and resemble the asymmetric sharp-cutoffs around the center frequency seen in cochlear filters in psychoacoustics and physiology \cite{glasberg1990derivation, bowman1998estimating}. 

This is followed by power-law compression, a lateral inhibition step highlighting change across frequencies, and short-term integration. These mimic processes in the cochlea and auditory nerve. 
In compression, we used $y = x^{\alpha}$ where $\alpha$ is a learnable parameter initialized at 1.0, and $x$ and $y$ the input and output. 
Lateral inhibition was done via a linear filter (initialized at $[1 -1]$) along the frequency followed by ReLU, which detects positive changes. For short-term integration, a leaky integrator with output $y(t) = e^{-t/\tau} x(t)$, with time constant $\tau$ was used. The output is downsampled to 200 Hz. 

The resulting auditory spectrogram differs from the conventional STFT spectrogram as it approximates a wavelet transform of the signal, better matches human auditory processing \cite{chi2005multiresolution}, and preserves spectrotemporal details for cortical analyses \cite{shamma2013balance}. Further, parameter choices of this model have been shown to improve performance in different listening environments \cite{shamma2013balance}, making the model suitable as a differentiable frontend, where task-specific parameters are learned. 
\subsection{Step 2. Cortical Features}
In the cortical step, joint spectrotemporal modulation features are extracted via filters from the auditory spectrogram. Each filter is characterized by spectral ($\Omega$) and temporal modulations ($\omega$). The spectrogram pattern that maximally excites the cortical spectrotemporal filter (or the STRF) roughly takes the shape of a Gabor kernel parameterized by $\Omega$ and $\omega$. The activation $r_f(t)$ for a single cortical neuron, tuned around a frequency $f$, at time instant $t$ can be computed as a function of frequency, scale, and rate by convolving the input auditory spectrogram $s(f,t)$ with the STRF kernel: 
\begin{align}
    r_f(t) = \text{STRF}(f, t; \Omega, \omega) *_{f,t}s(f,t)
\end{align}
where $\text{STRF}(.)$ stands for the 2D modulation bandpass filter.

This model is based on neuroscience experiments that explored transformations from the auditory spectrogram to cortical representations \cite{David2007-bm}, where modulation filters were designed to capture the important features in the cortical neural response. In this way, these filters decompose modulations of the auditory spectrogram into different pass-bands over a range of spectrotemporal modulations. This closely resembles the scattering transform \cite{Mallat2012-cs, Bruna2013-ba}. 
\subsection{Making the Model Differentiable}
We implemented the pipeline from waveform to cortical representations in a fully differentiable way using JAX \cite{jax2018github}. Our implementation is conceptually based on \cite{chi2005multiresolution}, adding vectorization to utilize GPU speedup. Additionally, all IIR filters were implemented in the Fourier domain, which is more tractable for gradient calculation than recursive filtering in the time domain. This allows parameters to be updated through back-propagation, using training schemes typical for neural networks such as minibatching and optimizers.

We initialize parameters with values usual in auditory models, and update them jointly with backend neural network parameters through backpropagation. The values for the poles and zeros of the constant-$Q$ filterbank are taken from \cite{chi2005multiresolution}. We initialized all values to 1.0 for power-law compression, $[1 -1]$ for the lateral inhibition filter, and $\tau=8$ for short-term integration. 
In the cortical stage, we initialized 40 spectrotemporal filters with both classical values (``log-spaced'', spaced logarithmically with more values at low spectral and temporal modulations) as well as uniform random initialization (``random'' within the range of $(0,9]$ in spectral and temporal modulations). Through training, we allowed all values to be freely updated except for the parameters in the constant-$Q$ filterbank, since allowing these updates would generate too many degrees of freedom. 

Our choices balance the adherence of the model to biological principles and simplicity. As prior work \cite{meng2023what} showed that cochlear filterbank parameters differed little from the Mel initialization when they are made learnable, we opt for a more complex and biologically plausible filterbank structure and do not update its values. In this way, it adheres closely to neural processes and offers more stable training and interpretability. For compression, the power-law compression only costs one parameter per frequency channel, and is a simplified version of PCEN \cite{wang2017trainable}. For lateral inhibition and short-term integration, we share parameters across channels, covering all channels with 3 parameters. We chose to focus the learnable parameters at the compression step based on previous work showing its relative importance in audio frontends \cite{meng2023what}. Our frontend has just 212 learnable parameters, and can be tweaked to increase or reduce parameters as needed.
\section{Applications of the Differentiable Frontend}
We show results on two types of tasks: classification (phoneme recognition) and enhancement (speech enhancement). We chose these as: {\em first}, both tasks are of interest to both technologists and neuroscientists, which means we can interpret the parameters using theories and human data (see \S \ref{sec:interp}); {\em second}, instead of limiting our front-end to classification like \cite{zeghidour2021leaf}, we also apply it to speech enhancement (an end-to-end problem). While some approaches (e.g., Conv-TasNet \cite{luo2019conv}) used learnable frontends, we show the benefits of constraining them to biologically plausible operations.

Cortical features that extract different spectrotemporal modulations from the acoustic signal could improve recognition of phonetic categories \cite{mesgarani2010multistream} and separate speech from various types of noise \cite{mesgarani2007denoising}, based on the hypothesis that different phonemes and speech sources generate distinct spectrotemporal modulation profiles, and thus are separable in the cortical representation. However, such feature-engineering is now less favored as they are outperformed by end-to-end methods. Here, we show that our differentiable model achieves comparable and often better performance compared with larger data-driven models. Additionally, the differentiable model shows superb robustness compared with their end-to-end counterparts. 
\subsection{Phoneme Recognition}
We test the differentiable frontend on phoneme recognition, where sound categories are predicted from the speech signal. While accuracy in phoneme recognition often positively correlates with ASR performance \cite{oh2021hierarchical}, phoneme recognition allows testing on small models without needing a language model.
\begin{figure}[htb!]
\centering
\includegraphics[width=\textwidth]{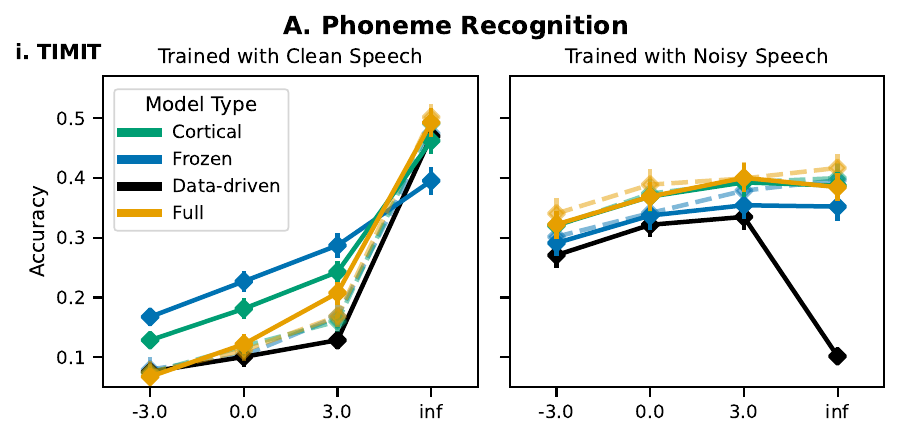}
\includegraphics[width=\textwidth]{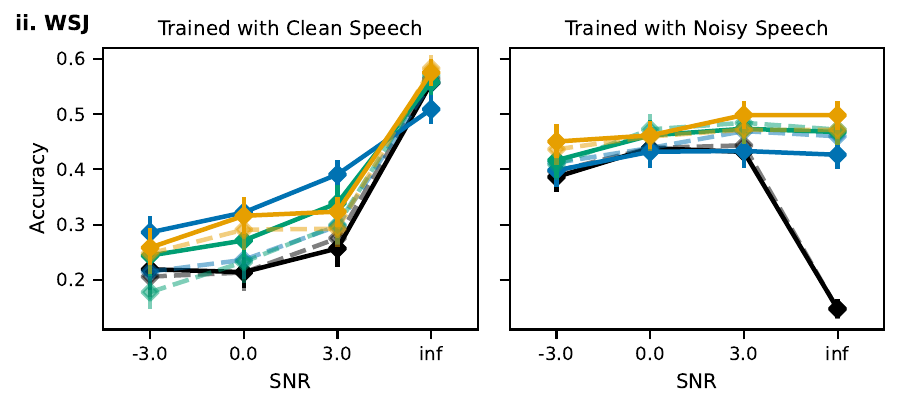}
\caption{Results for Phoneme Recognition. The $x$-axis shows test conditions where test speech was mixed with pink noise at -3, 0, and 3 dB SNR or clean. The $y$-axis shows classification accuracy. Solid lines denote models initialized with random cortical parameters, and dashed lines denote models initialized with log-spaced cortical parameters. Error bars denote 95\% CI.}
\label{fig:pr} 
\end{figure}

Phoneme recognition with our differentiable frontend takes in waveform input and outputs the spectrotemporal modulation values, which are fed into a 3-layer CNN, followed by a linear layer projecting to the number of phonemes (14.6k total parameters). The CNN layers have 3x3 filters with 10, 20, and 40 channels, each followed by GeLU. All parameters are learned jointly using the Adam optimizer, initial learning rate of 0.001, a batch size of 4, and for 200k steps. 
We obtained these parameters through a small grid search, although results are qualitatively similar with other hyperparameters. 

{\bf Ablation-test:}\ In the \textit{CNN} model, we replace the cortical step with an extra 3$\times$3 convolution layer with 40 channels, increasing the number of parameters since each cortical filter contains two parameters, while each CNN filter contains 10. This architecture performs feature extraction in a more data-driven way since the filters are not bound to bandpass the modulations. 
In the \textit{Frozen} model, both cochlear and cortical parameters are frozen during training and not updated, resembling the classical feature-engineering approach. In the \textit{Cortical} model, only cortical parameters were updated.

The model was trained and tested on two corpora, TIMIT \cite{garofolo1993darpa} and WSJ \cite{paul1992design} (with phone labels generated from forced alignment \cite{povey2011}). The training data for TIMIT consists of 4620 utterances, and WSJ was randomly subset to 4940 utterances to keep the duration comparable.
We trained the model on a modest GPU (a single RTX2080ti with 5 GB RAM), needing up to 12 hours of training time. In each training sample, one second of speech is randomly sampled from the training data. All models are tested on 100+ utterances of hold-out data (test split in TIMIT and randomly sampled for WSJ). 

Test accuracy is shown in Fig.~\ref{fig:pr}. We highlight two observations. First, differentiable models (Full and Cortical) outperformed theirs data-driven counterparts (CNN) in most cases. This shows the advantage of differentiability in performance and robustness, and aligns with observations in classical \cite{mesgarani2010multistream} and differentiable \cite{vuong2020learnable} modulation-based models. Second, we noticed that choices in parameter initialization (log-spaced vs random) influenced models' performance. When trained with clean speech, the randomly-initialized models generalized better to noisy conditions. This is supported by the distribution of the learned cortical filters, which we will elaborate in Section \ref{sec:interp}.
\subsection{Speech Enhancement}
We test our model on speech enhancement in a low-resource setting. We select this task to show that differentiable frontends need not only be used for classification (as they were in \cite{vuong2020learnable, zeghidour2021leaf}), but also in end-to-end tasks. Also, while separation and enhancement algorithms are often trained with big models and data (e.g. \cite{scheibler2023diffusion}), real-world hearing-aid/headphone processing scenarios require small models and data due to deployment constraints for personalized denoising. Here, we consider a task where speech needs to be separated from music. This setting is particularly relevant to our architecture since modulations are informative about auditory objects~\cite{elhilali2008cocktail, gu2008single, ding2017temporal}.

\begin{table}[b!]
\caption{SI-SDR for Speech Enhancement, higher is better. Best models in each row are bolded. 95\% CI are shown in brackets. }
{\tabcolsep=0.2cm \begin{tabular}{|l|ll|ll|ll|l|}
\hline
               & \multicolumn{2}{c|}{Full}                   & \multicolumn{2}{c|}{Cortical} & \multicolumn{2}{c|}{Frozen} & \multicolumn{1}{c|}{CNN} \\ \hline
Initialization & Log                  & Random               & Log           & Random        & Log          & Random       & -                       \\ \hline
Original       & 8.28(.24)          & \textbf{8.31(.23)} & 8.05(.24)   & 7.83(.2)   & 7.80(.24)  & 7.60(.2)  & 7.91(.25)\\
NewTarg.     & \textbf{5.85(.19)} & 5.09(.2)          & 5.30(.19)   & 5.09(.2)   & 5.22(.19)  & 4.57(.2)  & 5.39(.2)\\
NewNoise      & \textbf{16.2(.19)} & 15.5(.2)          & 16.0(.2)   & 15.6(.2)   & 15.2(.19)  & 15.0(.2)  & 16.1(.21) \\  \hline
\end{tabular}}
\label{ss}
\end{table}
The model uses the differentiable front-end followed by a CNN. Instead of classification labels, this CNN outputs a mask to be multiplied element-wise with the complex input STFT spectrogram (with window length of $256$ and hop length $80$), which is transformed back to waveform domain. The CNN has four layers with 20, 40, 10, 1 channels, with GeLU after each layer. The CNN outputs to a fully connected layer that projects the 129 frequency channels to the 256 STFT frequency channels, followed by sigmoid activation, to generate masks. We use the sum of L1 waveform loss and L1 multi-scale complex STFT spectrogram loss at window lengths [256, 512, 1024] and hop lengths at $1/4$ of the window length as loss function. For training, we used two hours of speech (from one female speaker, WSJ S002, 1000 utterances) and music (\cite{musdb18-hq}, 50 songs). In each training sample, 1 second of speech and music was randomly selected and mixed at 0 dB SNR. Other hyperparameters are identical to phoneme recognition. At test time, we evaluated the models on holdout data, as well as new distributions involving a new target speaker (WSJ S001, male) or a new type of noise (car noise from DEMAND \cite{thiemann2013demand}). We tested each condition with 500 samples mixed at 0dB SNR. 

The results (Table~\ref{ss}) show that the fully differentiable frontend was superior to all other ablations and data-driven counterparts. The log-spaced initialization significantly improved model performance. This is the opposite of the phoneme recognition results, which we discuss further below. 
\section{Explanability}
\label{sec:interp}
\begin{figure}[htb!]
\centering
\includegraphics[width=\textwidth]{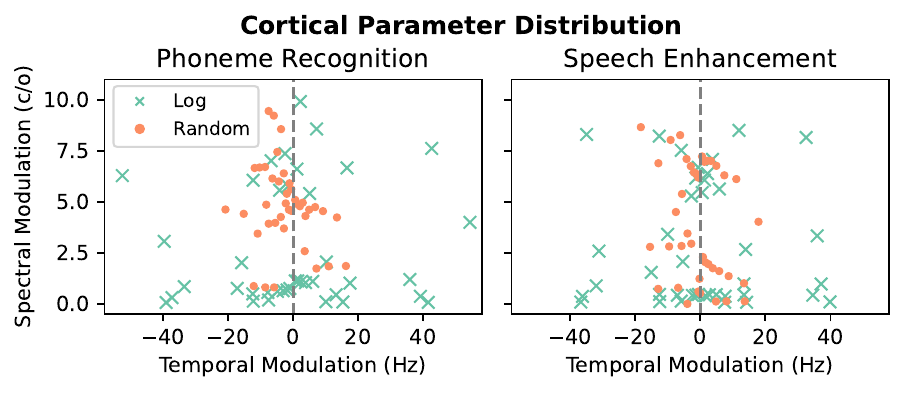}
\caption{Distribution of trained spectrotemporal filter parameters. Left: phoneme recognition in quiet; right: speech enhancement. Each point represents one of 40 filters. Two models are shown in each panel, respectively initialized from log-spaced values (cross) and randomized values (dot).
The sign of temporal modulation encodes the direction of the spectrotemporal modulation -- positive indicates upward-tilting Gabors and vice versa.} 
\label{fig:strf} 
\end{figure}

In addition to efficiency and robustness, since the differentiable front-end is based on neuroscientific processes, the parameters are directly interpretable. Here, we focus on the cortical parameters, while the full parameters including the learned cochlea parameters are published on our GitHub site. As the cortical stage band-passes in modulation domain, it serves as a bottleneck that discards modulations irrelevant to the task. Furthermore, the values of the parameter directly match different features in the audio: high spectral modulation ($>5$ cycles/octave) corresponds to narrow bandwidths related to spectral harmonics and pitch, while low spectral modulation ($<5$ cycles/octave) is related to spectral envelope information such as formants and timbre \cite{elliott2009modulation}. Thus, the learned parameters (shown in Fig.~\ref{fig:strf}) are indicative of how the model performs the task. 

The performance difference between two different initialization methods in the two tasks can be explained by the difference in the learned filter distributions. In Fig.~\ref{fig:strf}, the log-initialized models converged to filters spanning across a wider range of temporal modulations, up to $>50$ Hz, which are unrealistic for human cortical analysis or for modulation energy distribution in speech \cite{elliott2009modulation}. Performance-wise, we also noticed that the randomly initialized model, which converged towards realistic distributions, yielded better results. This observation aligns with prior differentiable modulation models in \cite{vuong2023incorporating}, that successful models converge to lower spectrotemporal modulations. However, in speech enhancement, the log-initialized models outperformed randomly initialized models. In the parameter distribution, the log-initialized model converged to temporal modulations within the biologically plausible range ($\approx 30$ Hz), and the higher temporal modulations likely helped separate speech from noise effectively through finer temporal detail. We conclude that the learned output of such differentiable models are sensitive to initialization and worthy of further study. Our results can also inform specific initialization choices. 

Furthermore, the learned distribution is closely related to the nature of the task. In the speech enhancement model, spectral modulations were contained within a lower range ($<8$ cycles/octave), which we attribute to the model learning the higher pitch of the target female speaker with lower spectral modulations.

\section{Discussion and Conclusions}
We present a differentiable front-end model of auditory processing combining signal processing, neuroscience, and deep learning, and apply it to example classification and enhancement tasks. Advantages of the differentiable model approach include the following. \textit{{\bf 1:}} Our differentiable frontend is lightweight (212 parameters total) and can be adapted to tasks with only a few hours of training data. \textit{{\bf 2:}} Compared with non-differentiable and data-driven counterparts, the differentiable frontend achieved better and generalizable performance \textit{{\bf 3:}} Model parameters are interpretable. As such, this frontend can be used for low-resource settings such as personalized denoising, where only minutes or a few hours of user data can be collected. Our model can also be applied to cases requiring high interpretability such as fitting hearing aid parameters. Additionally, since the model only employs linear filters, it can be adapted to be fully convolutional for arbitrary speech durations on modest hardware. 

The current model presents a middle ground between non-differentiable, spectrogram-based models and fully data-driven, end-to-end frontends (e.g., \cite{luo2019conv, baevski2020wav2vec}). Among similar cochlear \cite{zeghidour2021leaf, lyon2024carfac} and cortical \cite{vuong2023incorporating} models, our approach is rooted in auditory neuroscience, employing a biologically plausible filterbank and joint learning of cochlear and cortical stages, making it a better candidate for interpretability and hearing aid fitting. Also, our model does not employ any recursion, making back-propagation fast and stable. 
Lastly, the parameters we made differentiable were informed by prior work \cite{meng2023what}, pruning hundreds of parameters that deviate little after initialization. 

The model can be further optimized in the implementation of forward model and gradient calculations. In the current implementation, we focus on differentiability, but our model is fully compatible with CPU-based processing in deployment by simply changing the hand-engineered parameters with those learned through training. For this purpose, we have published Python implementation based on NumPy and SciPy on our Github. 
Also, gradient calculations could benefit from custom gradients to improve time and memory efficiency, where AD increases the memory footprint in an uncontrolled fashion for novel or complex functions.

The differentiable model can also be applied to hearing loss measurements and audio personalization. The cochlear parameters (e.g., compression) are directly related to hearing loss in different frequencies, allowing hearing loss to be characterized from user-collected data. Additionally, our forward model that predicts cortical responses can be adapted to be fitted with brain recordings, some of which are indicative of hearing loss through only a few electrodes \cite{shinn2017individual}. This allows us to fit parameters from brain recordings instead of cochlear measurements \cite{drakopoulos2023neural}, allowing for more degrees of freedom. This would lead to new types of supervised listening device fitting. 
\section{Acknowledgement}
We would like to thank Malcolm Slaney, Mounya Elhilali, and Dick Lyon for valuable discussions and feedback.

\end{document}